\documentclass[letterpaper,english,reprint,aps,superscriptaddress]{revtex4-1}

\usepackage{amsmath}
\usepackage{amssymb}
\usepackage{soul}
\usepackage{float}
\usepackage{graphicx}
\usepackage{subfig}
\usepackage{colortbl}
\usepackage[citecolor=blue,linkcolor=blue]{hyperref}
\usepackage{tikz}
\usepackage{lipsum, babel}

\newcommand{\ket}[1]{\ensuremath{\left|#1\right\rangle}}

\begin{document}

\title{Noise robustness and experimental demonstration of a quantum generative adversarial network for continuous distributions.}

\author{Abhinav Anand}
\email[E-mail:]{abhinav.anand@mail.utoronto.ca}
\affiliation{Chemical Physics Theory Group, Department of Chemistry, University of Toronto, 80  St, George Street, Toronto, Ontario M5S 3H6 Canada.}
\author{Jonathan Romero}
\affiliation{Zapata Computing Inc., 100 Federal St, 20th Floor, Boston, MA, 02110 USA}
\author{Matthias Degroote}
\affiliation{Chemical Physics Theory Group, Department of Chemistry, University of Toronto, 80  St, George Street, Toronto, Ontario M5S 3H6 Canada.}
\affiliation{Department of Computer Science, University of Toronto, 40 St. George Street, Toronto, Ontario, M5S 2E4, Canada.}
\author{Al\'{a}n Aspuru-Guzik}
\email[E-mail:]{aspuru@utoronto.ca}
\affiliation{Chemical Physics Theory Group, Department of Chemistry, University of Toronto, 80  St, George Street, Toronto, Ontario M5S 3H6 Canada.}
\affiliation{Department of Computer Science, University of Toronto, 40 St. George Street, Toronto, Ontario, M5S 2E4, Canada.}
\affiliation{CIFAR AI chair, Vector Institute, 661 University Ave. Suite 710, Toronto, Ontario, M5G  1M1, Canada.}
\affiliation{Canadian Institute for Advanced Research (CIFAR) Lebovic  Fellow, 661 University Ave, Toronto, ON M5G 1M1, Canada.}

\begin{abstract}
The potential advantage of machine learning in quantum computers is a topic of intense discussion in the literature. Theoretical, numerical and experimental explorations will most likely be required to understand its power. There has been different algorithms proposed to exploit the probabilistic nature of variational quantum circuits for generative modelling. In this paper, we employ a hybrid architecture for quantum generative adversarial networks (QGANs) and study their robustness in the presence of noise. We devise a simple way of adding different types of noise to the quantum generator circuit, and numerically simulate the noisy hybrid quantum generative adversarial networks (HQGANs) to learn continuous probability distributions, and show that the performance of HQGANs remain unaffected. We also investigate the effect of different parameters on the training time to reduce the computational scaling of the algorithm and simplify its deployment on a quantum computer. We then perform the training on Rigetti's Aspen-4-2Q-A quantum processing unit, and present the results from the training. Our results pave the way for experimental exploration of different quantum machine learning algorithms on noisy intermediate scale quantum devices.
\end{abstract}

\maketitle

\section{Introduction}

Quantum computers are expected to provide advantage over classical machines in certain sampling tasks \cite{boixo2018characterizing, aaronson2016complexity} because of their underlying quantum correlations, which could be helpful in modelling hard probability distributions. This has spurred much interest in investigating the possibility of achieving quantum advantage in quantum machine learning \cite{biamonte2017quantum}, leading to a lot of different quantum algorithms being proposed in the the last few years. However, the lack of perfect control in currently available quantum devices \cite{preskill2018quantum} limits the implementation of these algorithms to proof-of-principle experiments. The hybrid quantum-classical (HQC) \cite{mcclean2017hybrid, mcclean2016theory, zhu2019training} approach provides a way around this, by using the quantum resources in tandem with classical computers to improve the overall efficiency of the algorithm. 

In the last few years, the HQC approach has been frequently used to develop quantum algorithms for noisy intermediate scale quantum (NISQ) devices. Most of these algorithms use parameterized quantum circuits as physical ansatzes or statistical models, which are optimized by minimizing a cost function. Some examples include variational autoencoders (VAE) \cite{romero2017quantum, wan2017quantum, pepper2019experimental}, variational quantum eigensolvers (VQE) \cite{mcclean2016theory,peruzzo2014variational}, the quantum approximate optimization algorithm (QAOA) \cite{farhi2014quantum}, quantum generative adversarial networks (QGANs) \cite{romero2019variational, lloyd2018quantum, dallaire2018quantum,zoufal2019quantum, zeng2019learning, hu2019quantum}, among others.\cite{perez2020data, schuld2020circuit, grant2018hierarchical, bharti2021noisy, cerezo2020variational}

In the last couple of years, variational quantum circuits have been employed for generative modelling, particularly as GANs. Classical GANs have been used extensively for generative modelling and have become a very powerful tool in the machine learning community, for a variety of tasks \cite{creswell2018generative}, including image and video generation \cite{radford2015unsupervised,tulyakov2018mocogan}, and materials discovery \cite{sanchez2018inverse, zhavoronkov2019deep, gomes2019artificial, kim2020generative}. The quantum machine learning community has shown a great interest in exploring the use of variational quantum circuits to boost the performance of the classical GANs, because of their inherent ease of sampling data from quantum distributions. Many of the previous studies have explored this question for different scenarios \cite{romero2019variational}, and include proof-of-principle simulations and experimental demonstrations. One of these demonstrations focused on quantum state estimation \cite{hu2019quantum}, while the other learnt distributions by loading them in quantum states \cite{zoufal2019quantum}. While these results are very impressive, there is still need for extensive investigation about the performance of QGANs in presence of noise and the potential advantages of using them over their classical counterparts. 

\begin{figure*}[htbp]
  \centerline{
      \includegraphics[width=0.99\textwidth]{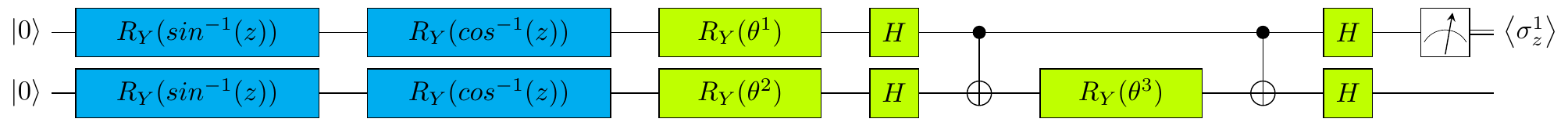}
    }
  \caption{The circuit architecture for the generator used in the simulation. The blue and green color represent the encoding and variational part of the circuit respectively.}
  \label{fig:gen_cir}
\end{figure*}

Various studies have explored the robustness to noise of different HQC algorithms such as VQE \cite{kandala2017hardware,barkoutsos2020improving} and QAOA \cite{dong2019robust}, where they show that these algorithms are to some extent resilient to it. In this work, we investigate the effect of noise on variational quantum circuits for generative modelling. We use the HQGAN model recently proposed by our research group \cite{romero2019variational} for learning classical probability distributions, and investigate its robustness with respect to noise. More specifically, we run simulations with realistic parameters to understand the effect of gate noise and other hardware imperfections on the model, before implementing it on a quantum computer.

The rest of the paper is organised as follows: In Section \ref{sec:HQGAN_theory}, we describe the theory of hybrid quantum GANs (HQGANs), the numerical simulation setup, and results. Section \ref{sec:Exp_implementation} describes the implementation of HQGANs on Rigetti's Aspen-4-2Q-A quantum processing unit, and presents the results obtained. Finally, a discussion of the significance of the experiment and the conclusion are presented in Section \ref{sec:conc}.

\section{Theory of Hybrid Quantum Generative Adversarial Networks}\label{sec:HQGAN_theory}

A prototypical GAN \cite{goodfellow2014generative} consists of two networks - a generator, $F_G(z;\theta_g)$ and a discriminator, $F_D(x;\theta_d)$ - playing an adversarial game, which can be summarized as follows:
\begin{equation}
\begin{split}
    \min_{\theta_g} \max_{\theta_d} (E_{x\sim p_{data}(x)}[log (F_D(x)] \\
    + E_{z\sim p_z(z)}[log(1-F_D(F_G(z)))]
\end{split}
\end{equation}

where $\theta_g$ and $\theta_d$ are the parameters of the generator and discriminator respectively, $p_z(z)$ is a fixed prior distribution for the generator to sample from and translate to samples that are indistinguishable from the real distribution $p_{data}(x)$, $x$ is the data sampled from the real distribution $p_{data}(x)$, and $z$ is the noise sampled from the prior distribution $p_z(z)$. 

The training of a GAN is carried out in an iterative manner where the discriminator and generator loss functions (equation \ref{eq:2} and \ref{eq:3}) are optimized alternatively, and the parameters of the discriminator and the generator are adjusted after each round of the optimization.
\begin{equation}\label{eq:2}
\begin{split}
   C_D(\theta_g, \theta_d) &=  E_{x\sim p_{data}(x)}[log (F_D(x)] \\
    &+ E_{z\sim p_z(z)}[log(1-F_D(F_G(z)))]
\end{split}
\end{equation}
\begin{equation}\label{eq:3}
    C_G(\theta_g, \theta_d) =  - E_{z\sim p_z(z)}[log(F_D(F_G(z)))]
\end{equation}
\\
The classical GAN uses neural networks (NNs) as discriminator and generators, to learn the target data distribution. However, a QGAN can be used in various scenarios - a fully quantum GAN with quantum circuits as discriminator and generator, for quantum or classical data; hybrid quantum GAN with quantum circuit as discriminator and NNs as generator, for quantum or classical data; hybrid quantum GAN with NNs as discriminator and quantum circuit as generator, for quantum or classical data. In this work, we employ a HQGAN that learns a classical target data distribution using quantum resources. We continue describing the details of the different quantum and classical networks we have employed in this work.

\subsection{HQGAN architecture}
The HQGAN conserves the two-component architecture of a regular GAN. While the discriminator used here is classical in nature, the generator uses operations on quantum states to perform its function. The generator is a two qubit quantum circuit, consisting of an encoding element and a variational element. The encoding element is built up out of two layers of single qubit rotation gates and uses a tensorial mapping strategy to introduce non-linearities \cite{mitarai2018quantum, schuld2019quantum}. The variational element consists of parametrized rotations and entangling (CNOT) gates. The angles in the rotations correspond to the parameters $\theta_g$ which are optimized during training. A circuit diagram of the generator is shown in Figure \ref{fig:gen_cir}. The data $x_{Fake} = \langle \sigma_z^1 \rangle$, is generated by measuring the expectation value of the first qubit in the $\sigma_z$ basis, and the reader is redirected to \cite{romero2019variational} for a detailed theoretical description of the generator.

The discriminator is a classical feed-forward neural network, with four layers - an input layer, two hidden layers with $50$ units each, and an output layer. The two fully-connected hidden layers ($1$ to $50$, and $50$ to $50$) have an exponential linear unit (ELU) activation function, and the final fully connected layer ($50$ to $1$) has a sigmoid activation function.

\subsection{Simulation setup}\label{sec:setup}
    
The HQGAN training was carried out by implementing the variational circuits using PyQuil \cite{smith2016practical}. The functions for expectation value and gradient calculation was added employing the autograd function from the PyTorch library \cite{paszke2019pytorch}, which enables us to do gradient based optimization. We use the Adam optimizer \cite{kingma2014adam}, and one-side label smoothing \cite{salimans2016improved} was used for both simulations and the experiment, which is a typical strategy used in classical GANs to improve convergence. The prior distribution is chosen to be a uniform distribution in the range $[-1,1]$. he expectation value calculations were done by doing 1000 circuit evaluations, unless stated otherwise. During the simulation, the metrics (Kullback-Leibler (KL) divergence, discriminator and generator losses, norm of the gradients, mean and standard deviation) during the training were calculated using 100 input data points sampled from the prior distribution, unless stated otherwise.

\begin{figure}[htbp]
\centering
\includegraphics[width=0.75\columnwidth]{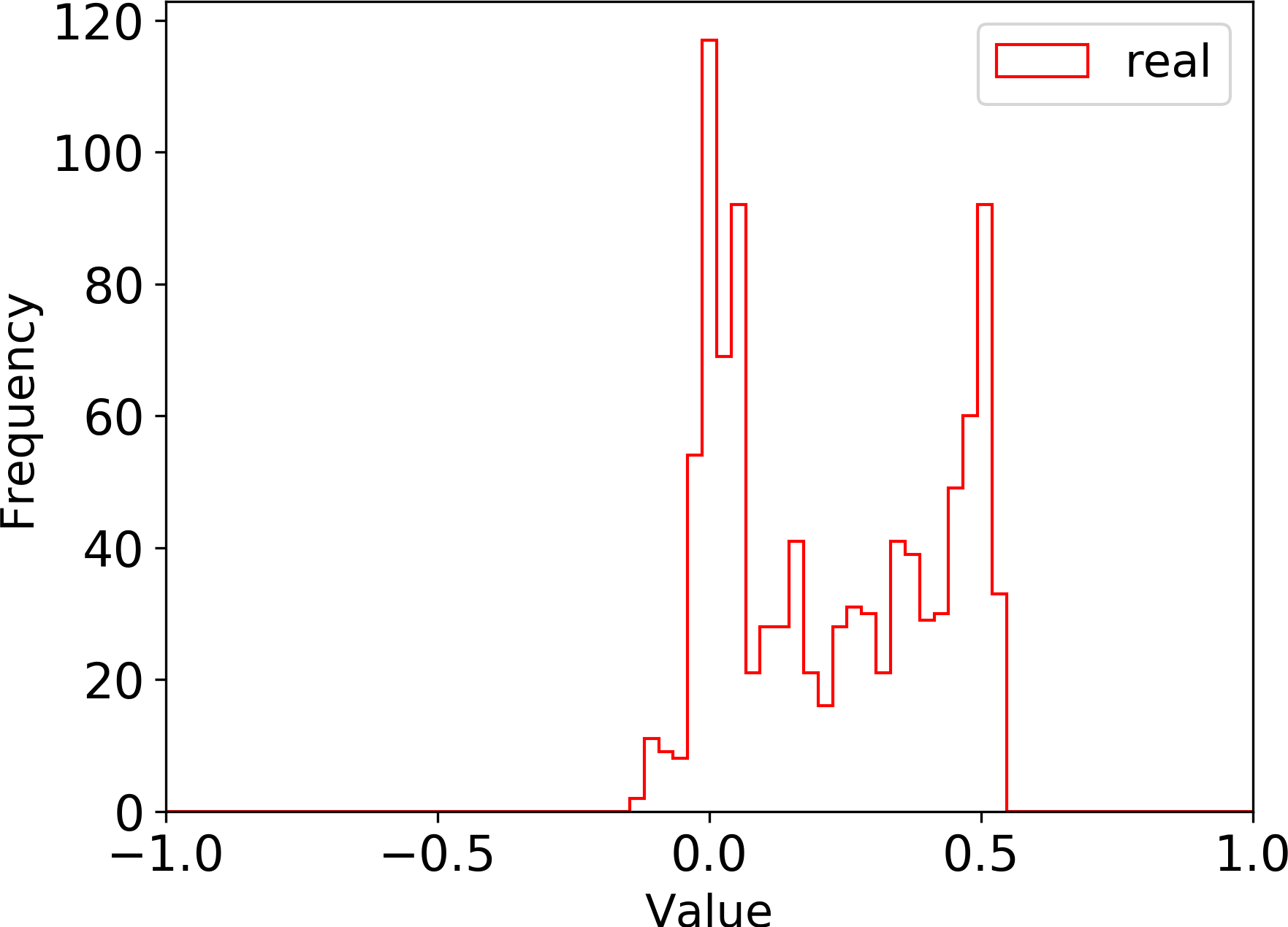}
\caption{The distribution of the target data generated by the variational quantum generator circuit (Figure \ref{fig:gen_cir}), with parameters fixed to $\theta_g = [ 0.35, 2.10, 5.06 ]$, using 1000 input data points from the prior distribution.}
\label{fig:Real_target}
\end{figure}

The target data for training was generated by measuring the expectation value of the first qubit by doing 1000 circuit evaluations of the quantum generator with the parameters fixed to $\theta_g = [ 0.35, 2.10, 5.06 ]$, and using 1000 input data points sampled from the prior distribution. The distribution of the target data generated by the generator is shown in Figure \ref{fig:Real_target}. We choose the initial parameters of the quantum generator, $\theta_g$ to be $[ 0.31, 1.89, 4.56 ]$ for all the simulations. The objective of this study is to investigate if QGANs can be used for generative modelling in the presence of noise, given the current state of the quantum resources. We chose initial parameters in the vicinity of the target parameters, as this would tell us the effect of noise on the functioning of the QGAN. The optimization and trainability of PQCs has been studied elsewhere \cite{wang2020noiseinduced}.

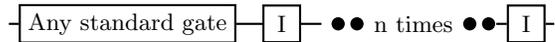
\begin{figure}[htbp]
  \centerline{
    \begin{tikzpicture}[thick]
    \tikzstyle{operator} = [draw,fill=white,minimum size=1.5em] 
    \tikzstyle{phase} = [fill,shape=circle,minimum size=5pt,inner sep=0pt]
    %
    \node at (-1.25,0) (q1) {};
    \node[operator] (op11) at (0.45,0) {Any standard gate} edge [-] (q1);
    %
    \node[operator] (op12) at (2.5,0) {I} edge [-] (op11);
    %
    \node (end1) at (3.15,0) {} edge [-] (op12);
    \node[phase] (phase) at (3.25,0) {} ;
    \node[phase] (phase1) at (3.5,0) {} ;
    \node (text) at (4.25,0) {n times};
    \node[phase] (phase3) at (5.0,0) {} ;
    \node[phase] (phase4) at (5.25,0) {} ;
    \node[operator] (op13) at (5.75,0) {I} edge [-] (phase4);
    \node (end2) at (6,0) {} edge [-] (op13);
    \end{tikzpicture}
  }
  \caption{A sample implementation of a noisy gate, the standard gate is any gate after which noise has to be added. The number n depends on the gate time, as follows: $n = t_{gate} / t_{I-gate}$.}
  \label{fig:noisy_gate}
\end{figure}

To evaluate the performance of HQGANs in realistic execution conditions, we have studied our model in the presence of noise sources that model the errors in quantum hardware. The noise was introduced in the simulations by adding noisy gates to the variational quantum circuit (the generator component of the GAN). Noisy identity gates (with a single Kraus map) were added after all the gates in the generator circuit. The number of such noisy gates, was decided based on the gate time, as follows:  $n = t_{gate} / t_{I-gate}$, where $t_{gate}$ is the single qubit or two qubit gate time, depending on the gate being single or two qubit, and $t_{I-gate}$ is the gate time for the noisy identity gate, which is equal to single qubit gate time (see Table \ref{tab:params} for details). A sample implementation of a noisy gate in the circuit is shown in Figure \ref{fig:noisy_gate}.

\begin{table*}[htbp]
    \centering
    \begin{tabular}{c|c}
        \hline
        Parameters & Values \\
        \hline
        \hline
        amplitude damping time $T_1$& 15 $\mu s$ \\ 
        dephasing time $T_2$ & 18 $\mu s$\\  
        one-qubit gates times $t_1$ & 50 $ns$  \\
        two-qubit gates times $t_2$ & 400 $ns$ \\
        readout assignment probabilities $p_{0|0}$, $p_{1|1}$ & 0.91 \\
        \hline
        \hline
        \hline
        \multicolumn{2}{c}{Probabilities of noise in simulations with single noise models} \\
        \hline
        Amplitude damping probability, $p$ & 0.09 \\
        Phase damping probability, $p$ & 0.09\\
        \hline
        \hline
        \hline
        \multicolumn{2}{c}{Probabilities of noise in all other simulations} \\
        \hline
        Amplitude damping probability, $p$ & $1 - \exp(-t/T_1)$\\
        Phase damping probability, $p$ & $1 - \exp[-2t(1/T_2 - 1/2T_1)]$ \\
        \hline
    \end{tabular}
    \caption{ A table containing all the parameters for different noise models used in the simulations. The expression for the two probabilities can be derived by examining the evolution of the density matrix with time.~\cite{Ghosh_2012}  The experimental values for $T_1$, $T_2$ and gate times $t_1$ and $t_2$ were taken from the Rigetti cloud services website~\cite{pyquilqcs} in the month of June, 2019. The probability values for single qubit gates based on the experimental parameters, for amplitude damping is $0.0033$ and for phase damping is $0.0022$. \textbf{NOTE}: The values of the probability for phase damping noise model has to be divided by 2, to be used in simulation with pyquil because of their different Kraus map implementation~\cite{pyquilnoise}. }
    
    \label{tab:params}
\end{table*}

We used different directive statements (pragmas) available in the Forest platform \cite{smith2016practical} to modify our noiseless circuit. The pragmas inform the quantum virtual machine (QVM) that a gate is to be replaced with an imperfect realization using a Kraus map in the noisy simulation. The noise models we use in our simulations include: 1) amplitude damping; 2) dephasing; 3) decoherence (a combination of amplitude damping and dephasing), and 4) a combination of amplitude damping, dephasing and readout noise. The Kraus operators corresponding to amplitude damping and dephasing are shown in equation \ref{eq:amp_damp} and \ref{eq:dephasing} below:
\begin{equation}\label{eq:amp_damp}
    K_1 = 
    \begin{pmatrix}
    1 & 0 \\
    0 & \sqrt{1-p}
    \end{pmatrix},
    \text{and }
    K_2 = 
    \begin{pmatrix}
    0 & \sqrt{p} \\
    0 & 0
    \end{pmatrix}
\end{equation}

\begin{equation}\label{eq:dephasing}
\begin{split}
    K_1 &= 
    \begin{pmatrix}
    \sqrt{1-p} & 0 \\
    0 & \sqrt{1-p}
    \end{pmatrix}, \\
    K_2 &= 
    \begin{pmatrix}
    \sqrt{p} & 0 \\
    0 & 0
    \end{pmatrix}
    \text{, and } \\
    K_3 &= 
    \begin{pmatrix}
    0 & 0 \\
    0 & \sqrt{p}
    \end{pmatrix}
\end{split}
\end{equation}

where, $p$ is the probability of the qubit decaying/dephasing over a time interval of interest. The operator $K_2$ for the amplitude damping noise model controls how the state decays from the state $\ket{1}$ to $\ket{0}$, while the operator $K_1$ describes the evolution of the state in the absence of a quantum jump. The evolution of the density matrix in presence of the amplitude noise model can be expressed as follows:
\begin{equation}\label{eq:ev_damp}
    S(\rho) = 
    \begin{pmatrix}
    \rho_{00} + p\rho_{11}  & \sqrt{1-p}\rho_{01} \\
    \sqrt{1-p}\rho_{10} & (1-p)\rho_{11}
    \end{pmatrix},
\end{equation}
 while the combined effect of the operators for the dephasing noise model, is the reduction of the transverse component of the density matrix, which can be expressed as:
\begin{equation}\label{eq:ev_deph}
    S(\rho) = 
    \begin{pmatrix}
    \rho_{00}  & (1-p)\rho_{01} \\
    (1-p)\rho_{10} & \rho_{11}
    \end{pmatrix},
\end{equation}
The decoherence noise model used in the simulations is a combination of the damping and dephasing noise model, and the Kraus operators for the combined noise model are obtained by combinatorially multiplying the operators of the two noise models. The readout noise model is modeled according to the assignment probability matrix, which has two independent parameters, $p_{0|0}$ and $p_{1|1}$, representing the conditional probability $p(x'|x)$, where $x'$ is the output when $x$ is transmitted through a noisy channel. The parameters for the noise models used in simulations are listed in Table \ref{tab:params}. We present here the simulations with single noise sources with a failure probability $p = 0.09$, which is very high compared to the probability calculated from experimental parameters. This value was chosen based on a grid search with simulations carried out to investigate the possibility of optimization at different levels of error. This was the highest value at which we were able to observe a reasonably well functioning HQGAN. Combining multiple noise sources at this high noise level results in very bad performance of the HQGAN. For that reason we show simulations for the combined noise channels with a lower noise level that more closely resembles the experimental parameters. A detailed description of the noise models used here can be found in Refs. \cite{ pyquilnoise,preskill2015chapter}.

As with classical computation, wall-time on a quantum device is a limited resource. We carried out simulations to estimate and reduce the time required to run the experiment on a quantum computer. We investigated the effect of two parameters, the number of data points sampled from the distributions for training, and number of circuit evaluations for expectation value calculation. The reason behind investigating these parameters is that they influence the total number of circuit evaluations that one has to perform for training. The first, the number of input samples affects the objective function calculation for the QGAN, while the second, the number of circuit evaluations for expectation value calculation affects the output of the quantum generator. To reduce the values of these parameters and hence the computational overhead, we investigate the dependence of the optimization on these parameters. We discuss the results of the simulations in detail in the next sub-section.

\begin{figure*}[t]
\centering
    \includegraphics[width=0.95\textwidth]{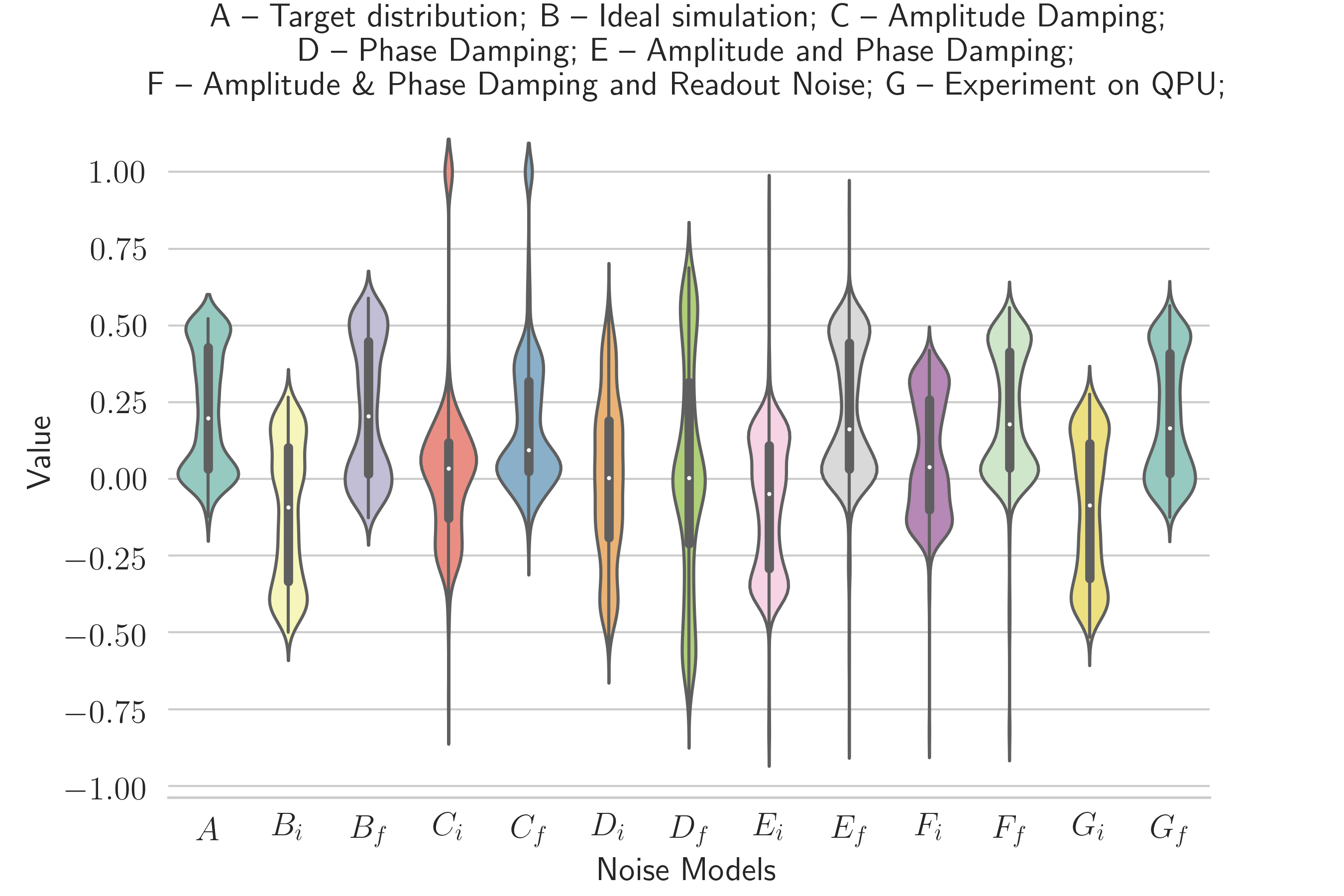}
    \caption{Violin plots illustrating the probability distributions obtained from an HQGANs experiment. Details of the model are described in the main text. Plots correspond to A target distribution, B simulated distribution generated in absence of noise. The next distributions correspond to simulation under different noise models: C amplitude damping ($p=0.09$), D dephasing ($p=0.09$), E amplitude damping ($p=0.0033$) and dephasing ($p=0.0022$) and F a combination of amplitude damping ($p=0.0033$), dephasing ($p=0.0022$) and readout noise. The final distributions correspond to G an experiment using the Aspen-4-2Q-A  2-qubit chip from Rigetti Computing. The labels $i$ and $f$ indicate the distribution before and after training respectively (see Section \ref{sec:sim_res} for details). The data for the plot was generated by sampling 1000 data points from the quantum generator with the optimized parameters after the training is completed.}
    \label{fig:simulation_res}
\end{figure*}

\subsection{Simulation results}\label{sec:sim_res}
The training was done over 4500 epochs. This cutoff value is chosen to do a fair comparison between the different HQGAN training carried in this study, and is larger than the average number of iterations required in the different simulations. The training is evaluated by tracking metrics such as the KL divergence of the two distributions, discriminator and generator losses, norm of the gradients, mean and standard deviation of the distributions. We visualize the data in violin plots to compare the distributions generated from the simulations. The plots can be interpreted as follows, the white dot in the middle represents the median, the thick line the interquartile range, the thin line the rest of the distribution except the outliers, and the area around the probability density using kernel density functions. The data for the plots are generated by using the optimized generator parameters with 1000 points sampled from the generator after training is finished. The metrics from various simulations are plotted in section II of the appendix.

\subsubsection{Noiseless simulations}       
We simulated an ideal generator circuit with $100$ input points $z$, sampled from the uniform prior distribution $U(-1,1)$, and 1000 circuit evaluations for every measurement value. The results from the simulations are plotted in Figure \ref{fig:simulation_res} and labeled $B_i$ and $B_f$, representing the distribution before and after the training. It can be seen from the plots that the distribution generated by the quantum generator at the end of 4500 epochs of training is an approximation of the target distribution, with the median, interquartile range and distributions resembling the corresponding metrics for the target distribution (labeled A) also shown in Figure \ref{fig:simulation_res}. This is in agreement to the results from the numerical simulations carried out in our earlier work\cite{romero2019variational}. We also carried out another ideal simulation with initial parameters equal to all zero, and present the results in section I of the appendix.

\subsubsection{Noisy simulations}
The noisy simulations were carried out with the same parameters as the noiseless simulation above. The distributions from the different simulations and the training metrics are plotted in Figure \ref{fig:simulation_res}. Different letters represent different noise models. The results show that under the presence of noise, all the simulations with the exception of the one under purely dephasing noise, approximately converged to the target distribution. The distributions for noisy simulations show visible tails products of noise.

First, we ran simulations to test the effect of the damping and dephasing noise models on the training of the HQGAN. We investigated the effect by using different damping/dephasing probabilities, and show the results from the training at probability $p$ = 0.09, which is a very high value compared to the error probability expected from the experiments.

Under the influence of purely amplitude damping noise with a damping probability $p$ = 0.09, (label C) we observe that a visible part of the distribution centers around expectation value 1.0, as a consequence of the noise driving the population of states to $\ket{0}$. The amplitude damping noise channel drives the average of the population towards an expectation value of zero, as observed from the distributions previous to training. 

However, when the training is successful, the generated distribution recovers the shape and moments of the target one despite the effect of noise, showing how the training of variational circuits is still possible under moderate noise conditions.

In the case of purely dephasing noise with a damping probability $p$ = 0.09, the initial and final distributions are considerably distorted compared to the noiseless case, becoming a symmetric distribution (label D) centered at zero, as seen in Figure \ref{fig:simulation_res}. The phase damping channel unlike the amplitude damping channel does not involve any transition, but leads to the system decaying to a incoherent superposition in the computational basis, which could be the reason for the spread in the distribution. However, after training, the final distribution gains some features that resemble the target distribution, such as the modes around 0.5 and 0.0, but still has a third spurious mode around -0.6.

After analyzing the training with a large damping/dephasing probability, we next ran simulation with different combinations of the noise models. However, we now use the parameters in Table \ref{tab:params}, to compute the probabilities according to the expressions, $p_{damping} = 1 - \exp(-t/T_1)$ and $p_{dephasing} = (1 - \exp[-2t(1/T_2 - 1/2T_1)])$.
The distribution generated by the combination of different noise models (label E and F), do not show a significant deviation from the ideal simulation, as the probability values are significantly smaller compared to the value of 0.09 employed in the previous simulations. 
It is also worth pointing out that our numerical experiments indicate that the presence of moderate noise facilitates convergence by reducing the number of epochs required. Improvements in optimization of other algorithms involving parameterized quantum circuits have been previously reported as well \cite{schuld2020circuit,grant2018hierarchical,huggins2019towards}.

\subsubsection{Input samples and Circuit evaluations}
\begin{figure}[h]
\centering
     \subfloat[A corresponds to the target distribution. Labels B, C, D and E correspond to distributions obtained using 25, 50, 75 and 100 input data points respectively during training. Labels $i$ and $f$ at the end indicate distribution at Epoch:0, and Epoch:2000 respectively.]{%
      \includegraphics[width=0.95\columnwidth]{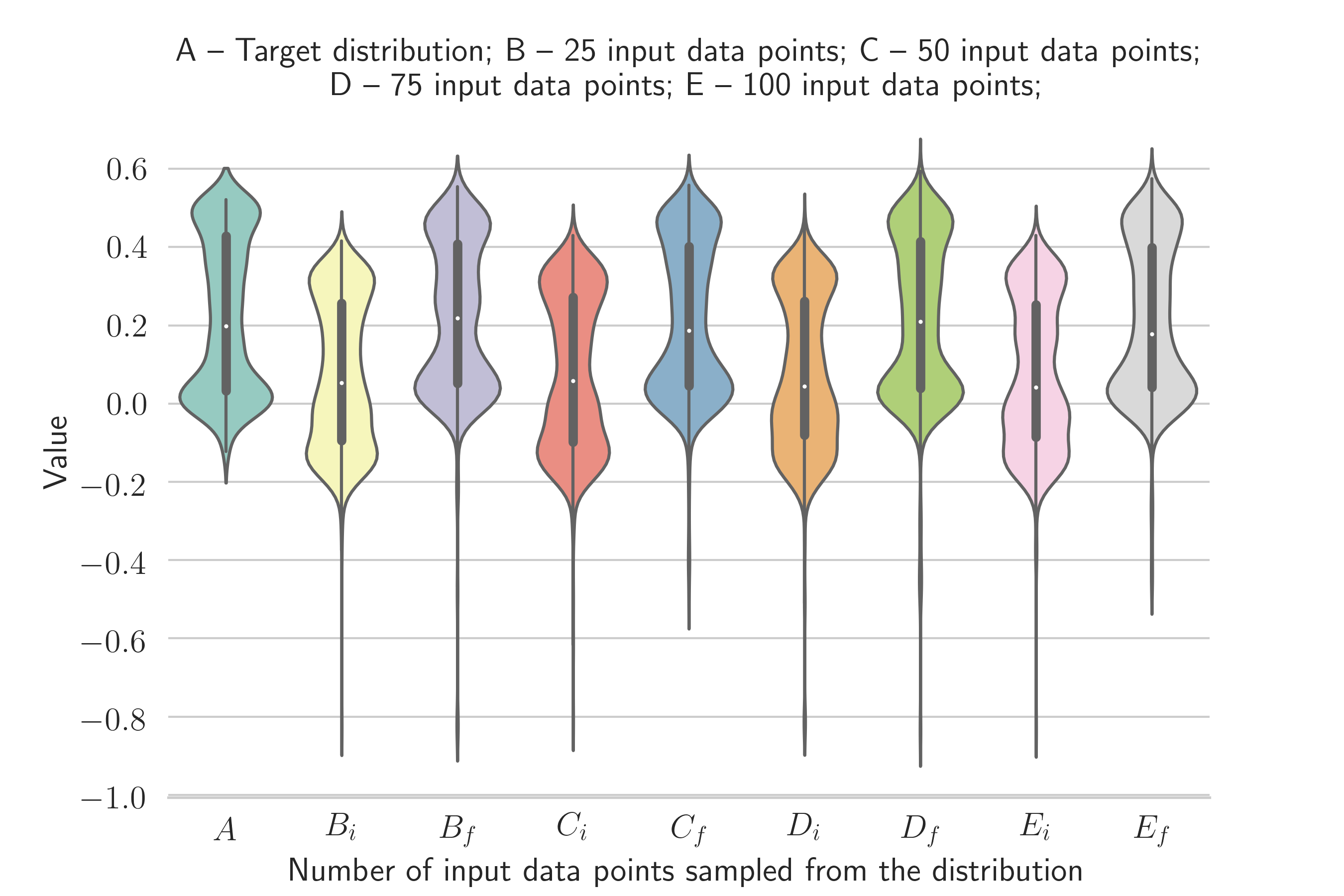}%
    }

    \subfloat[A corresponds to the target distribution. Labels B, C, D and E correspond to the distributions obtained with 100, 250, 500 and 1000 circuit evaluations respectively during training. Labels $i$ and $f$ at the end indicate distribution at Epoch:0, and Epoch:2000 respectively.]{%
      \includegraphics[width=0.95\columnwidth]{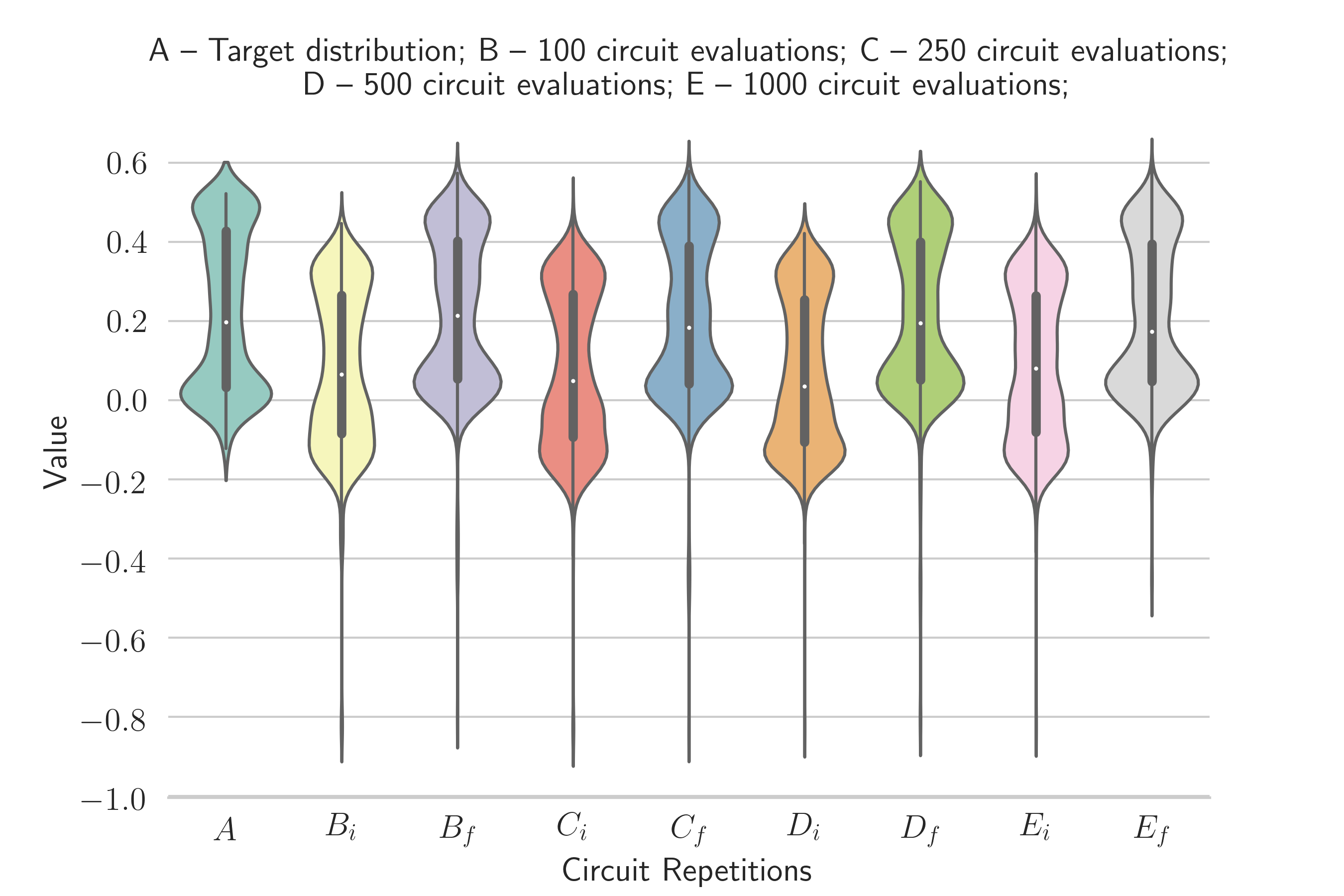}%
    }
 
   \caption{Violin plots comparing the effect of the number of input data points for training, and number of circuit evaluations used to compute expectation values in HQGANs. All the data points for the plot were generated by 1000 data points sampled from the quantum generator with the optimized parameters after the training is completed.} 
    \label{fig:exp_metrics}
\end{figure}

Next, we ran simulations to investigate the ability of the algorithm to converge as a function of the number of input samples used per epoch of the HQGAN training. Using less samples reduces the amount of computational resources required for training. We used four different values of the number of input samples, $25, 50, 75,$ and $100$, and the combination of amplitude damping, dephasing, and readout noise models for the simulations, with the parameters derived from the values in Table \ref{tab:params}. We plot the distributions generated from the training, in Figure \ref{fig:exp_metrics}a. We observe that the number of input samples used for every epoch of training has very little effect on the training, and the distributions converge to a good approximation of the target distribution in all the simulations after 2000 training epochs. 

After observing that reducing the number of samples does not seem to decrease the quality of the training, we studied the impact of varying number of measurement shots used to estimate expectation values. The number of input samples was fixed to 25 for all of these simulations, and we chose $100, 250, 500,$ and $1000$ as the different number of circuit evaluations for expectation value calculation. Simulations were carried out with the combination of the amplitude damping, dephasing, and readout noise models, with the same parameters as for previous simulations. 

The generated distributions are plotted in Figure \ref{fig:exp_metrics}b, and it is evident that the training is unaffected by the number of circuit runs. This is in agreement with the result from previous works \cite{sweke2019stochastic, harrow2019low}, where it was shown that for various hybrid quantum-classical optimization algorithms, the estimation of the expectation values can be done using very small number of measurements. The number of samples can be considered a hyper-parameter of the algorithm that could be tuned or adjusted during the calculation. Based on the simulations, we deduced that we could reduce the run-time of the experiment with a minimal effect on the accuracy by using smaller values for input samples and circuit evaluations.

\section{Implementation on Quantum Hardware}\label{sec:Exp_implementation}
In this Section, we present details on the implementation and execution of an HQGAN on the Aspen-4-2Q-A superconducting quantum processor from Rigetti computing \cite{Karalekas_2020}. We use the same target state, generator and  discriminator architecture as in the simulation. We used our findings from noisy simulations regarding the hyper-parameters for the training to optimize the run-time on the QPU. We set the number of samples drawn from the distribution for an epoch of the training to 25, and the number of circuit runs per evaluation of the expectations value to 250.

\begin{figure}[htbp] 
    \centering
        \subfloat[Distributions at Epoch 0.]{\includegraphics[width=0.95\columnwidth]{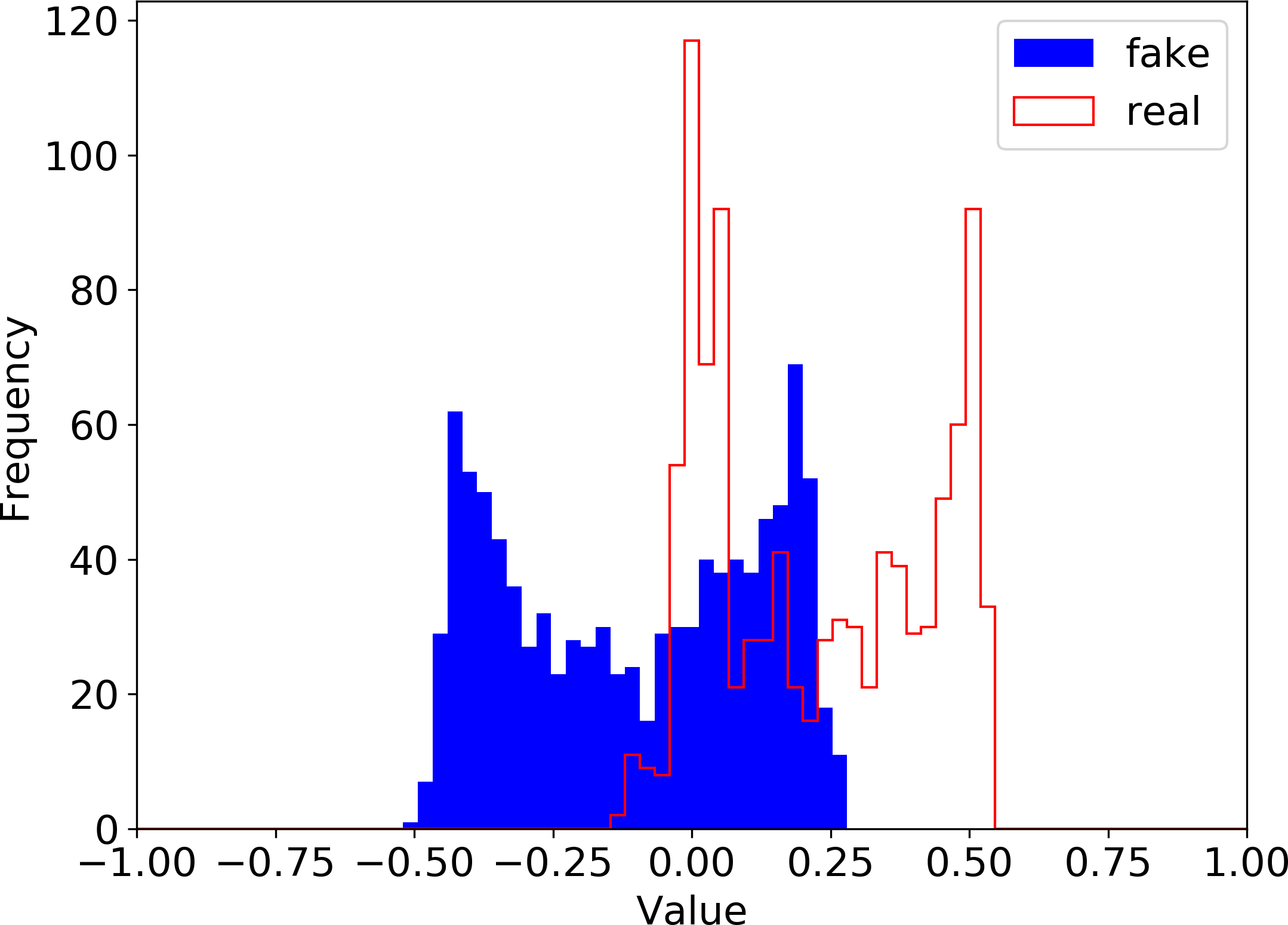}}
        
        \subfloat[Distributions at Epoch 1500.]{\includegraphics[width=0.95\columnwidth]{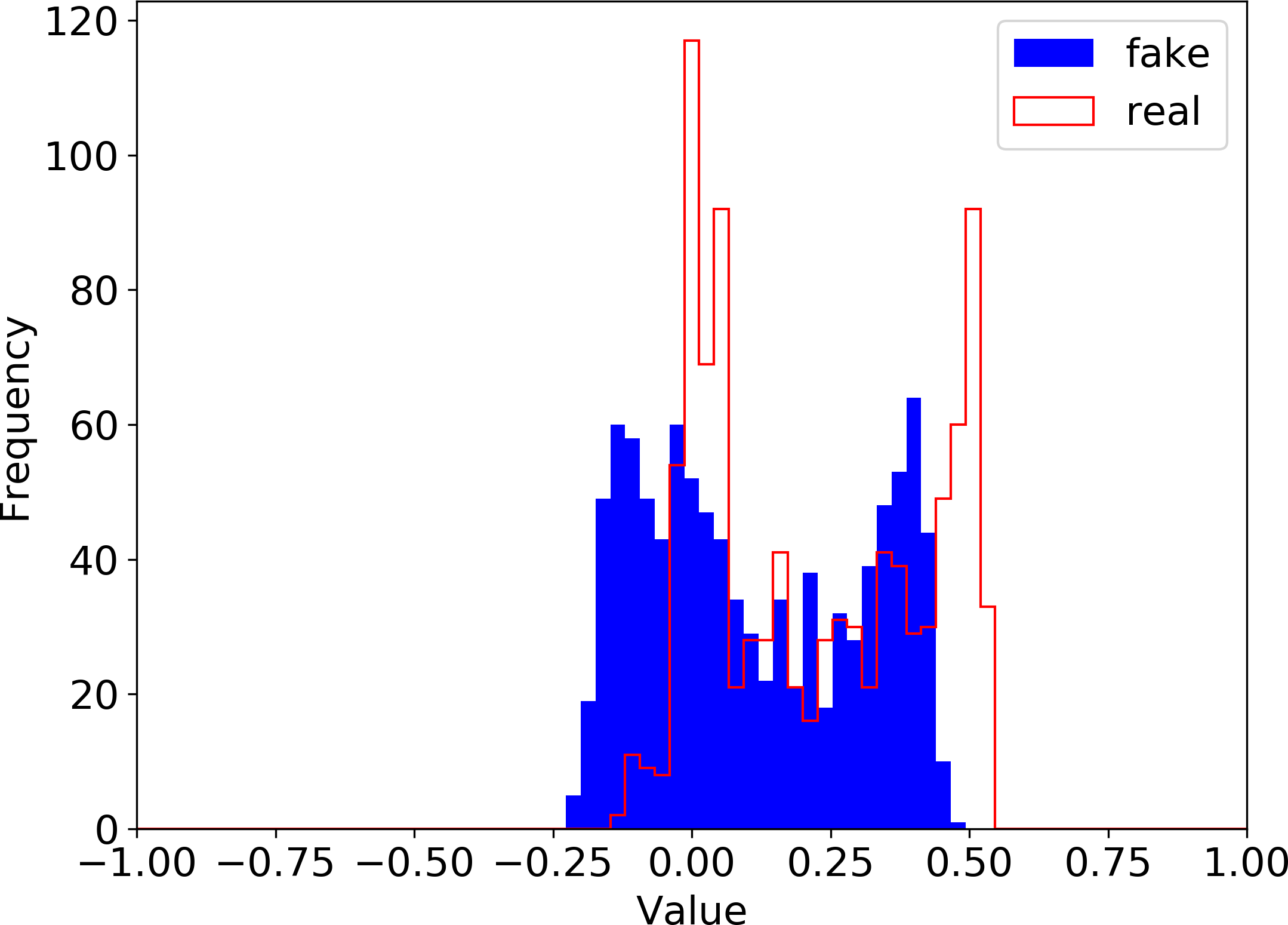}}
        
        \subfloat[Distributions at Epoch 2993.]{\includegraphics[width=0.95\columnwidth]{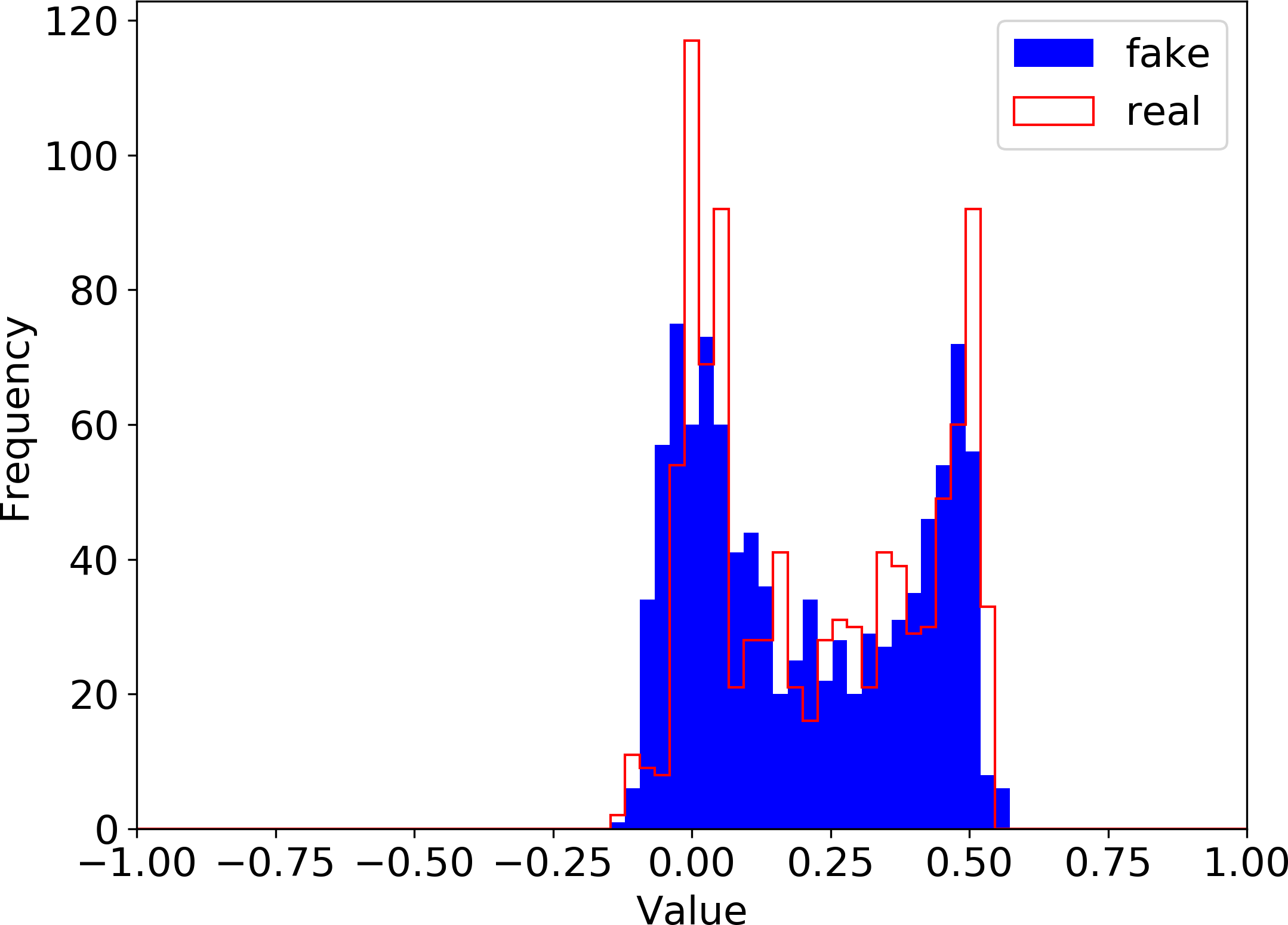}}
    
     \caption{Snapshots of the data distribution generated by the quantum generator at different epochs during the HQGANs training. The quantum generator corresponds to a parameterized 2-qubit quantum circuit executed on the Aspen-4-2Q-A superconducting quantum processor. The discriminator used is a classical neural network. All the distributions were generated using 1000 input points sampled from the prior distribution and then running the optimized generator circuit after the training is completed.}
    \label{fig:QPU_evolution}
\end{figure}

Based on the simulations and gate times, we estimated that the full training would require more than a day on the QPU. We divided the full experiment into two hour slots and completed the training by running $\sim30$ such slots. We finished $2993$ epochs of training, which was sufficient for the HQGAN to learn the target distribution. Figure \ref{fig:QPU_evolution} illustrates the evolution of the distribution generated by the generator at different epochs during the run on the quantum computer. Figure \ref{fig:QPU_evolution} shows how the final distribution from the quantum generator achieves a good overlap with the target distribution.

\begin{figure}[htbp]
\centering
 \includegraphics[width=0.95\columnwidth]{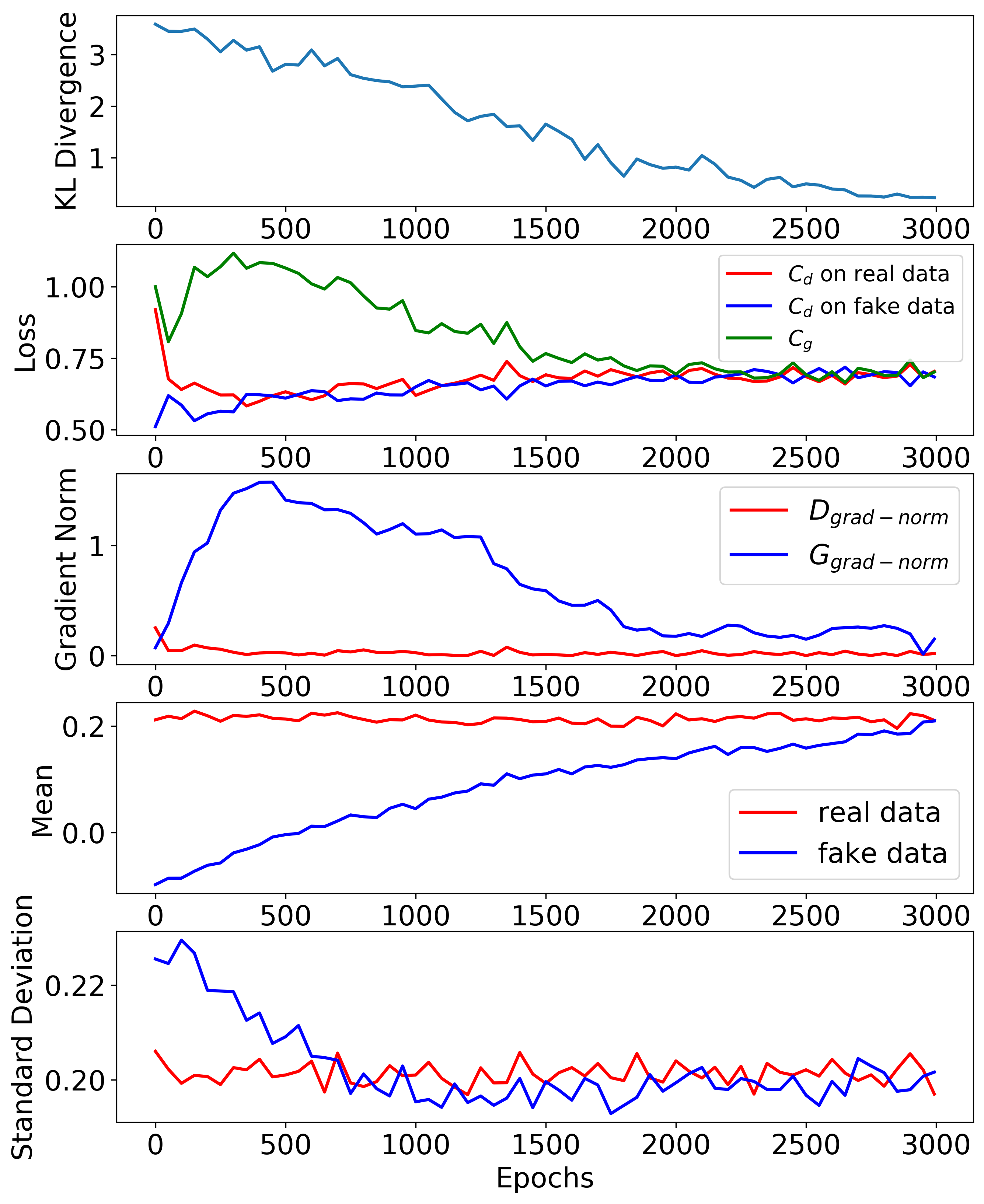}%
\caption{From top to bottom: Kullback-Leibler divergence between target and generated distribution, objective functions ($C_d$ and $C_g$), Norm of the gradients($D_{grad}$ and $G_{grad}$), mean and standard deviation of the two distribution during the training on the QPU as a function of the number of epochs. All the statistics were computed using 1000 data points sampled from the quantum generator with the optimized parameters after the training is completed.}
\label{fig:QPU_stats}
\end{figure}

To evaluate the training on the QPU we tracked the KL divergence and objective functions of the generator and discriminator. The top two plots in Figure \ref{fig:QPU_stats} show the dynamics of these metrics, indicating how the KL divergence decreases to nearly $0$ during the training and the cost functions converge to approximately a value of $ln(0.5) \approx 0.7$. We also plot the mean, standard deviations, and gradient norms of the distribution in Figure \ref{fig:QPU_stats}. The plot illustrates that the mean, standard deviations, and gradient norms converge for both distributions.

We also plot the distribution before and after the training, obtained from our experiment in Figure \ref{fig:simulation_res} (label G). It can be seen from the plots that the results from the simulation matches the distribution obtained in the  experiment. 
This can be attributed to the fact that the distributions and other statistics including, KL divergence, objective functions of the generator and discriminator, mean, standard deviations, and gradient norms of the distribution,  were computed using the 1000 data points sampled from the generator with the parameters from training carried out on the physical quantum device. 
Thus using the parameters from the optimization of the quantum GAN on the noisy quantum device as that of the generator, the cost functions and final distribution obtained from the sampled data is very similar to the target distribution and the final distribution obtained in the noiseless simulation.
This phenomenon has been observed in the case of variational quantum compiling~\cite{sharma2020noise}, where the authors termed this as optimal parameter resilience (OPR),
and show that the variational parameters remain unaffected by different types of incoherent noise. 
Our study further strengthens the belief that NISQ devices can be used for successfully optimizing variational algorithms.

This result confirms the ability of the procedure to succeed under moderate levels of noise, and adds to the growing practical evidence showing that algorithms that rely on optimizing a parameterized quantum circuit exhibit a strong resilience to the moderate level of noise on a NISQ device.

\section{Conclusion}\label{sec:conc}

In this work, we have demonstrated the training of a HQGAN \cite{romero2019variational} on a quantum computer. We evaluated the performance of our proposed HQGAN with respect to different noise models. We have ran simulations to reduce the computational scaling of the experiment, before performing the training on a quantum device. Our numerical demonstrations using both simulated and physical quantum devices show that the HQGAN training can be carried out in the presence of moderate levels of noise. We also found empirical evidence that we can perform the training with reduced number of input samples per epoch, and calculate expectation values for optimization with fewer circuit evaluations. We used the Aspen-4-2Q-A $2$-qubit chip from Rigetti Computing for performing the HQGAN training, obtaining results similar to those obtained from classical simulation.

Our numerical exploration illustrates how NISQ devices can be used for generative learning and contributes to the growing field of parameterized quantum circuits as machine learning models. Proof of principle demonstrations, such as those shown here, constitute a first step towards using quantum resources to enhance existing classical machine learning pipelines. One such direction is investigating the advantage of using the current protocol for practical tasks, such as image, speech and text generation. We will keep exploring different strategies to enhance the applicability of the current protocol to different scientific applications of broader interest.

\section{Acknowledgements}

Simulations were performed on the Niagara supercomputer at the SciNet HPC Consortium~\cite{Loken_2010,10.1145/3332186.3332195}. SciNet is funded by: the Canada Foundation for Innovation; the Government of Ontario; Ontario Research Fund - Research Excellence; and the University of Toronto. We acknowledge access to Rigetti quantum computer via their cloud computing services, and thank the team and Amy Brown in particular for their help with the QPU simulations. A.A and A.A.G. acknowledge support by the  U.S. office of naval research. M.D. and A.A.G acknowledge support by the U.S. Department of Energy, Office of Science, Office of Advanced Scientific Computing Research, Quantum Algorithms Teams Program. A.A.G. acknowledges the generous support from Google, Inc. in the form of a Google Focused Award.  A.A.G. is grateful for the funding from the Vannevar Bush Faculty Fellowship Program, and the Canada Research Chairs Program. A.A.G thanks Anders G. Fr\o{}seth for his generous support.

\bibliography{hqgan.bib}

\newpage

\section*{Appendix}

\subsection*{I. Ideal simulation}

\begin{figure}[htbp]
    \centering
    \includegraphics[width=0.95\columnwidth]{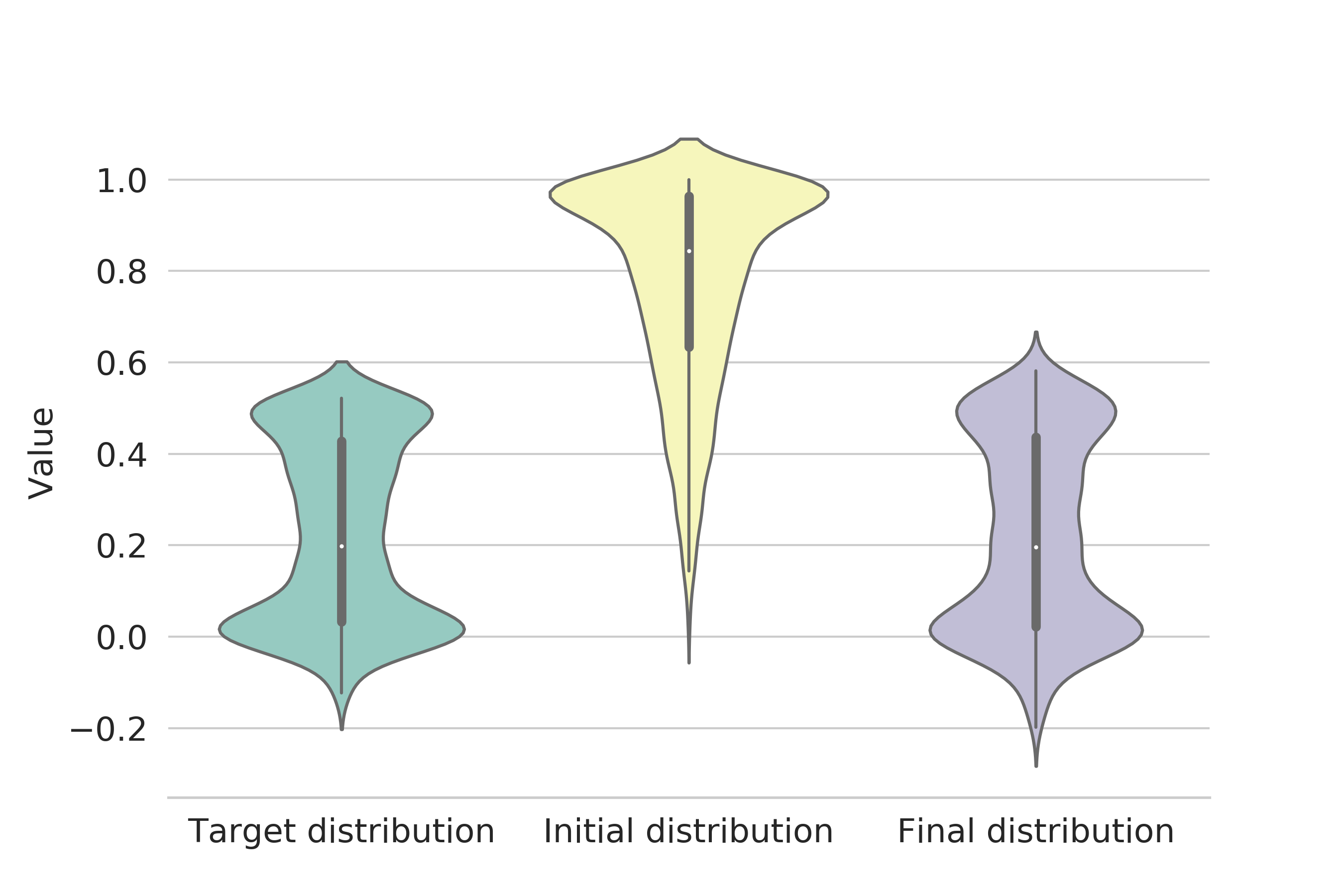}
    \caption{Violin plots showing the distribution of the target data and the sampled data from the generator at different epoch during the training. All the data points for the plot were generated by 1000 data points sampled from the quantum generator with the optimized parameters after the training is completed.}
    \label{fig:sim_dis_id_0}
\end{figure}

\begin{figure}[H]
    \centering
    \includegraphics[height=150pt]{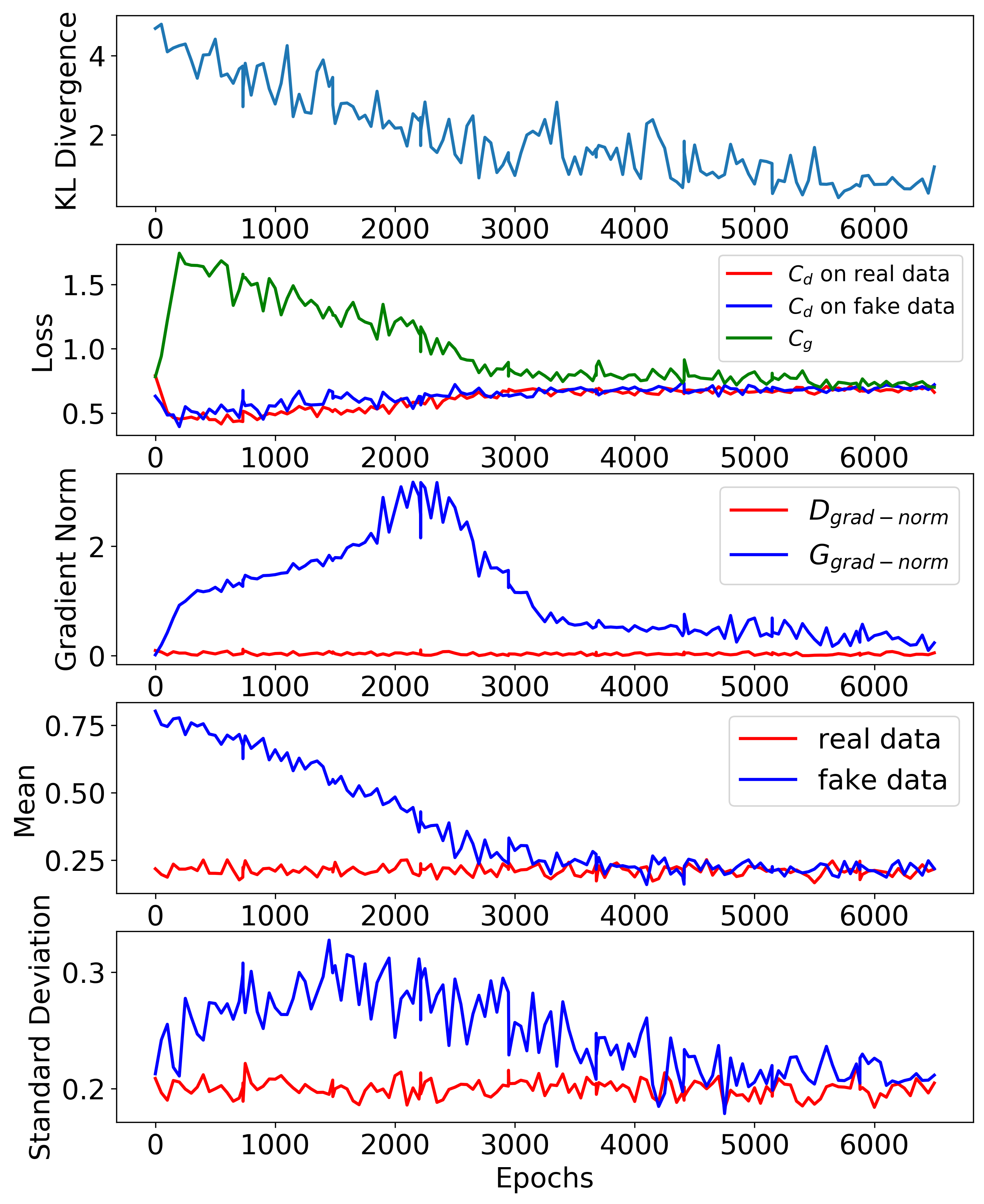}
    \caption{The metrics from the ideal simulation with the parameters of the generator initialized to zero. From top to bottom: Kullback-Leibler divergence between target and generated distribution, objective functions ($C_d$ and $C_g$), Norm of the gradients($D_{grad}$ and $G_{grad}$), mean and standard deviation of the two distribution during the training as a function of the number of epochs.All the statistics were computed using 100 data points sampled from the quantum generator with the optimized parameters after the training is completed.}
    \label{fig:sim_met_id_0}
\end{figure}

We carried out another simulation of the ideal generator circuit with $100$ input points $z$, sampled from the uniform prior distribution $U(-1,1)$, and 1000 circuit evaluations for every measurement value, but changed the initial value of $\theta_g = [ 0.0, 0.0, 0.0]$. The distribution before and after 6500 epochs of training is plotted using violin plot in Figure \ref{fig:sim_dis_id_0}, and metrics (Kullback-Leibler (KL) divergence, discriminator and generator losses, norm of the gradients, mean and standard deviation) in Figure \ref{fig:sim_met_id_0}. It can be seen from the plots that the distribution and the metrics converge to that of the target distribution, similar to the other simulations but require more rounds of training to achieve full convergence. 

\subsection*{II. Other simulation Metrics}
We plot the metrics from all the simulations with the different noise models here.

\begin{figure}[H]
    \centering
    \includegraphics[height=190pt]{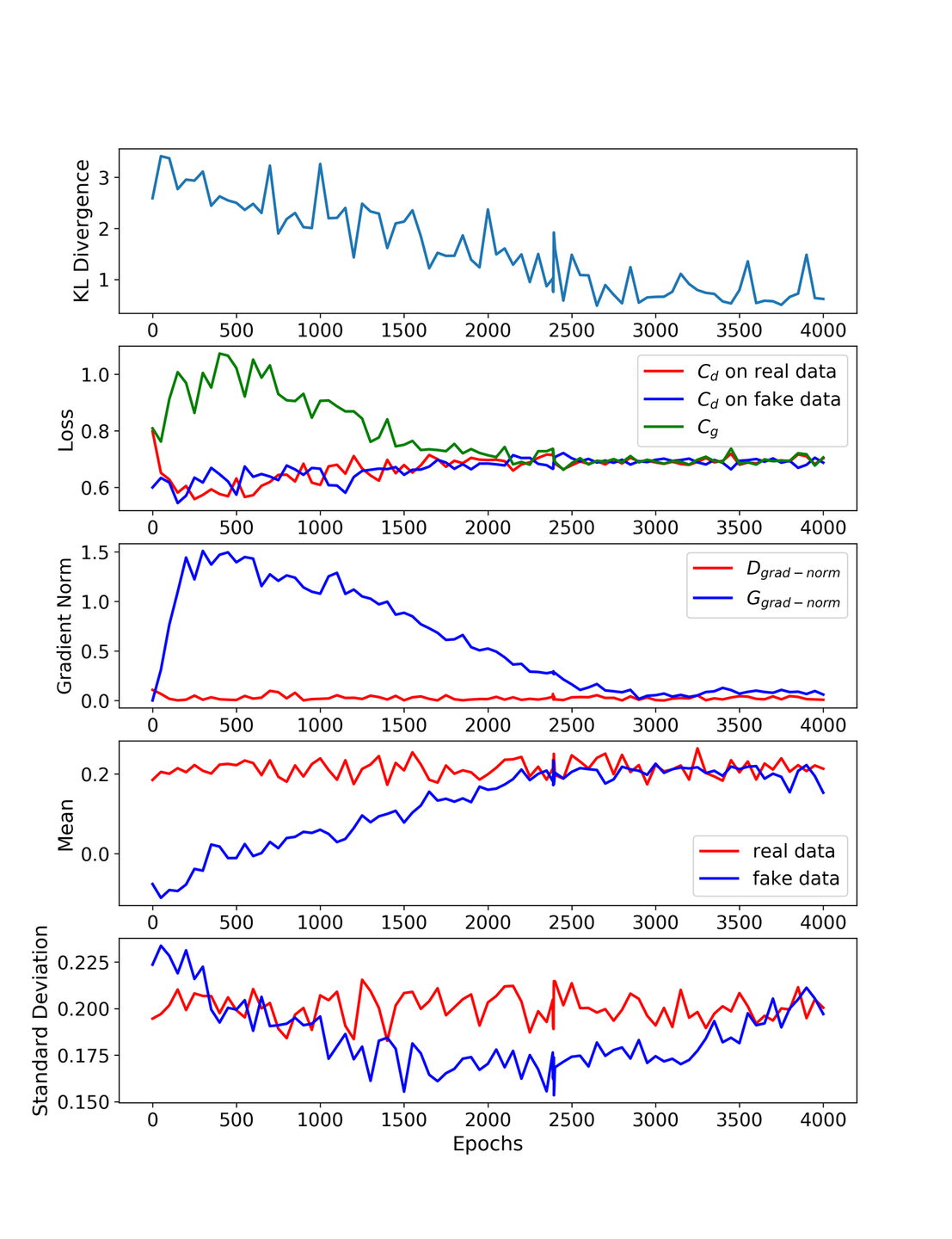}
    \caption{The metrics from the ideal simulation. From top to bottom: Kullback-Leibler divergence between target and generated distribution, objective functions ($C_d$ and $C_g$), Norm of the gradients($D_{grad}$ and $G_{grad}$), mean and standard deviation of the two distribution during the training as a function of the number of epochs.All the statistics were computed using 100 data points sampled from the quantum generator with the optimized parameters after the training is completed.}
    \label{fig:sim_met_none}
\end{figure}

\begin{figure}[H]
    \centering
    \includegraphics[height=190pt]{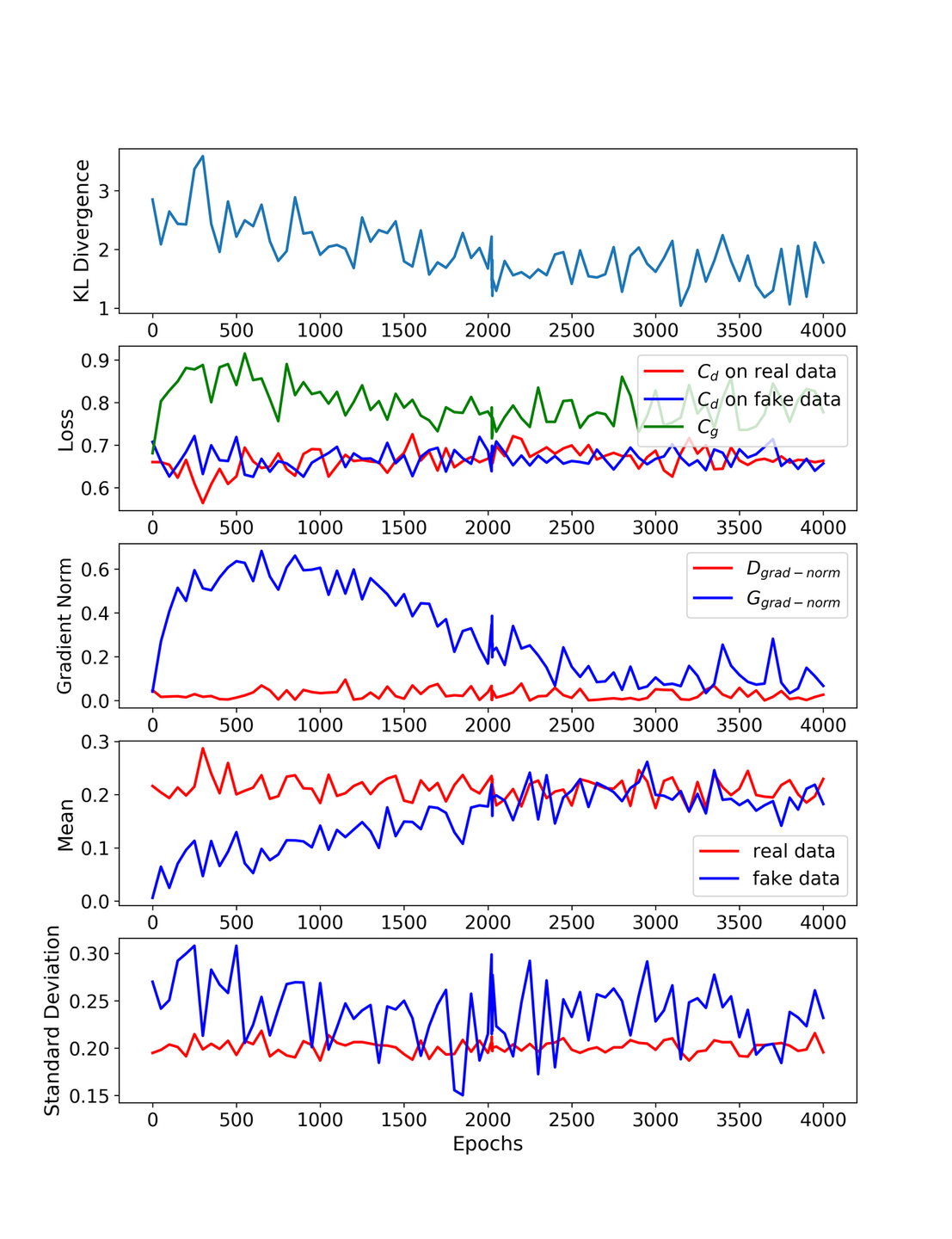}
    \caption{The metrics from the simulation with amplitude damping noise model only. From top to bottom: Kullback-Leibler divergence between target and generated distribution, objective functions ($C_d$ and $C_g$), Norm of the gradients($D_{grad}$ and $G_{grad}$), mean and standard deviation of the two distribution during the training as a function of the number of epochs.All the statistics were computed using 100 data points sampled from the quantum generator with the optimized parameters after the training is completed.}
    \label{fig:sim_met_amp}
\end{figure}

\begin{figure}[H]
    \centering
    \includegraphics[height=190pt]{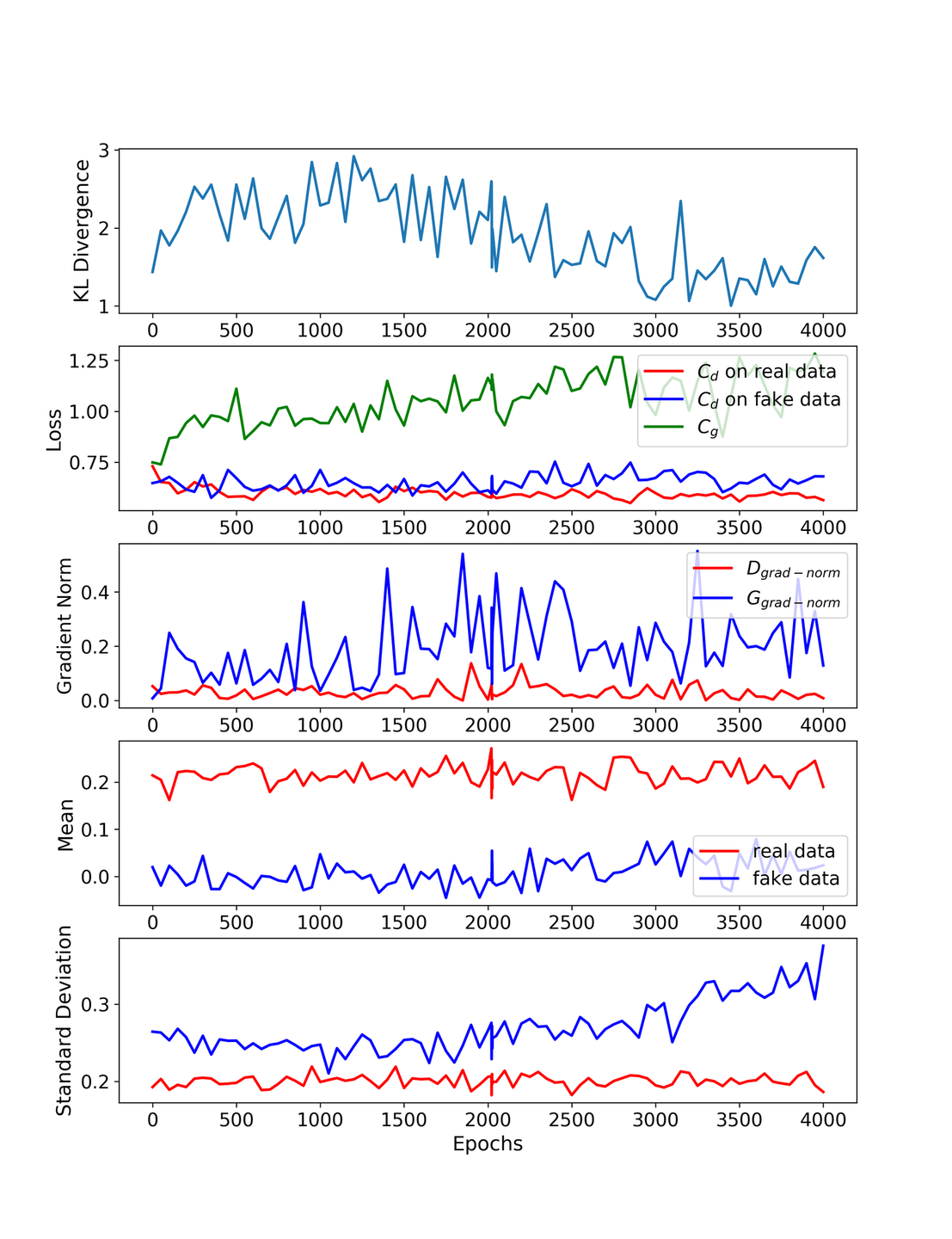}
    \caption{The metrics from the simulation with phase damping noise model only.From top to bottom: Kullback-Leibler divergence between target and generated distribution, objective functions ($C_d$ and $C_g$), Norm of the gradients($D_{grad}$ and $G_{grad}$), mean and standard deviation of the two distribution during the training as a function of the number of epochs.All the statistics were computed using 100 data points sampled from the quantum generator with the optimized parameters after the training is completed.}
    \label{fig:sim_met_phase}
\end{figure}

\begin{figure}[H]
    \centering
    \includegraphics[height=190pt]{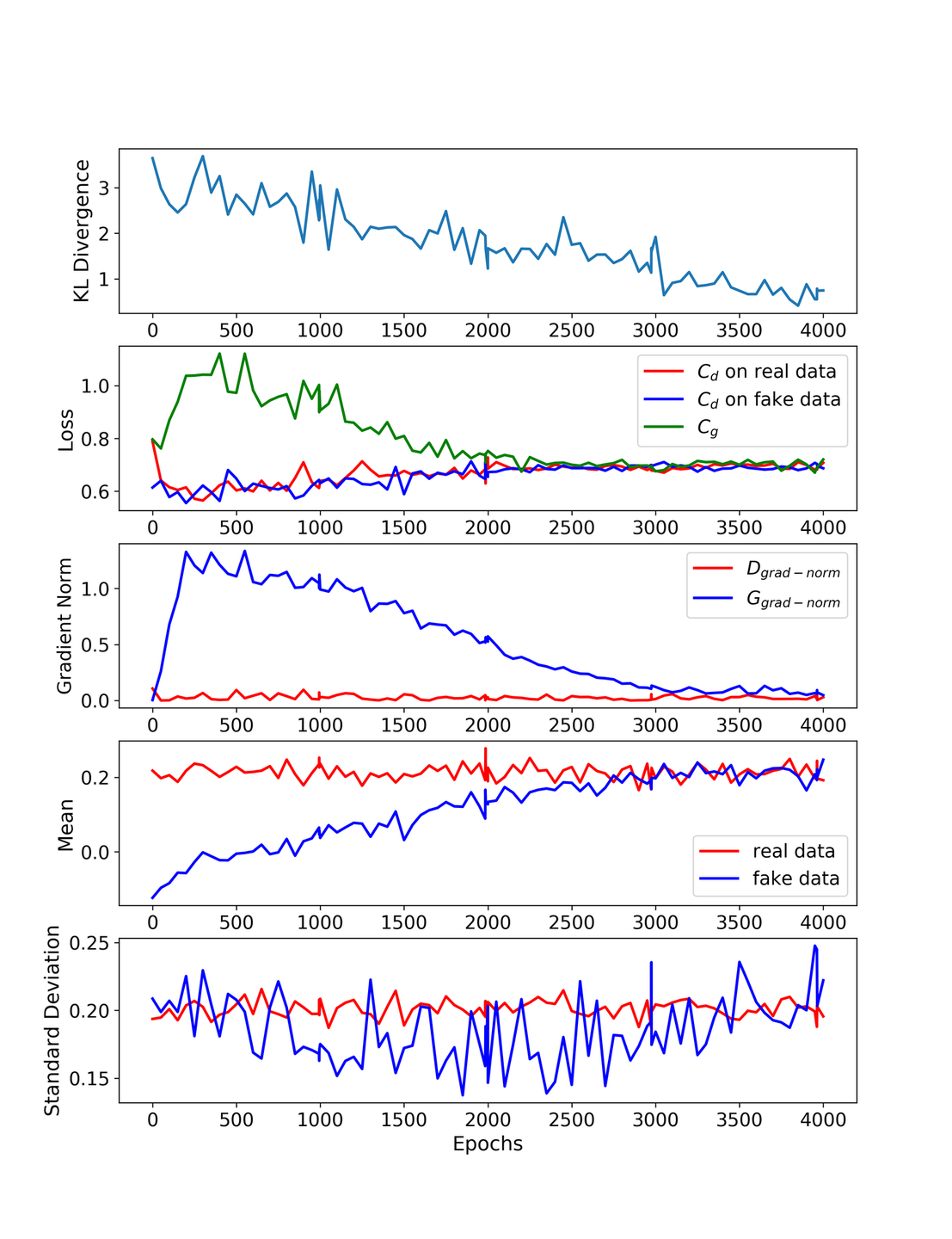}
    \caption{The metrics from the simulation with combined amplitude and phase damping noise models.From top to bottom: Kullback-Leibler divergence between target and generated distribution, objective functions ($C_d$ and $C_g$), Norm of the gradients($D_{grad}$ and $G_{grad}$), mean and standard deviation of the two distribution during the training as a function of the number of epochs.All the statistics were computed using 100 data points sampled from the quantum generator with the optimized parameters after the training is completed.}
    \label{fig:sim_met_dec}
\end{figure}

\begin{figure}[H]
    \centering
    \includegraphics[height=190pt]{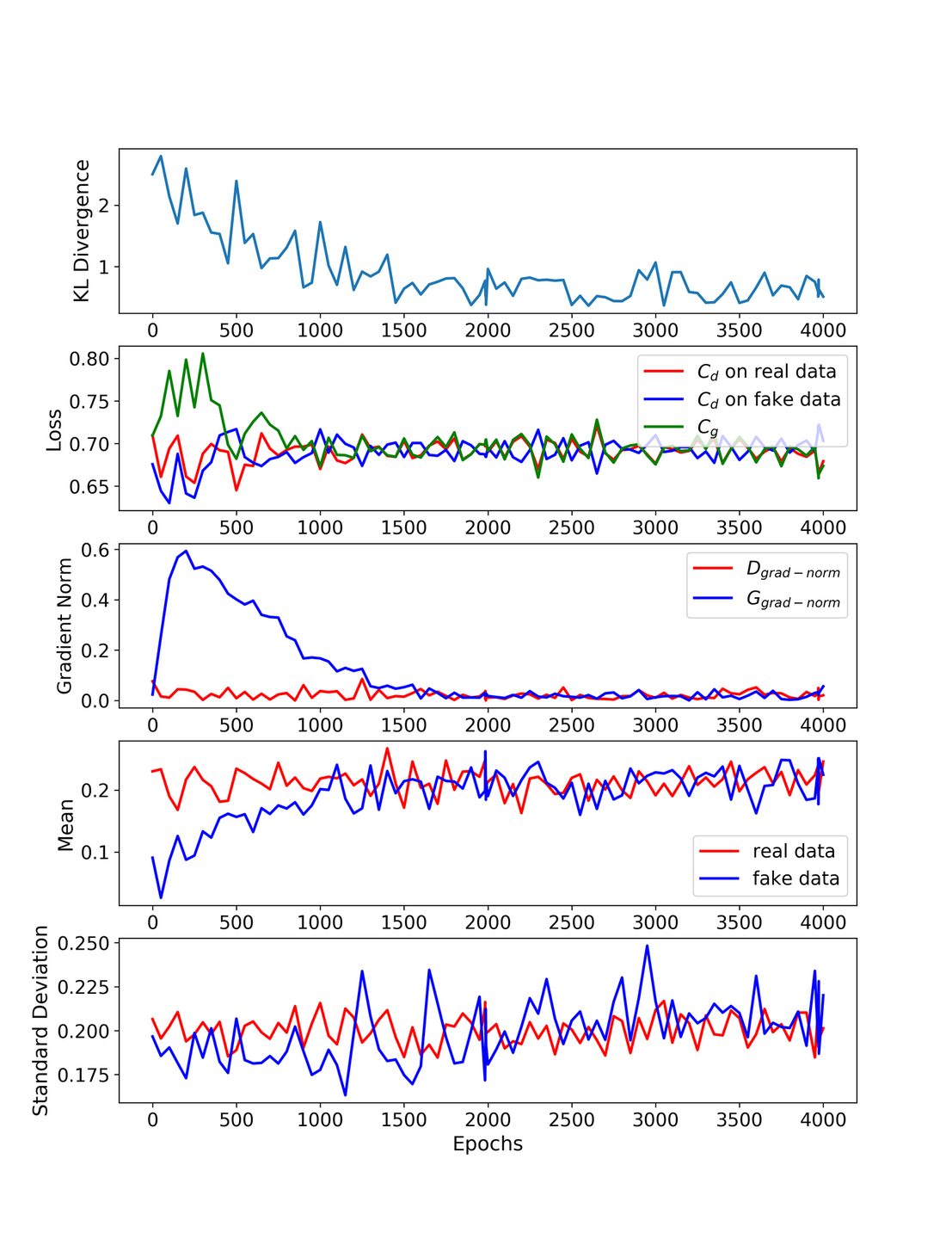}
    \caption{The metrics from the simulation with a combination amplitude \& phase damping and readout noise models.From top to bottom: Kullback-Leibler divergence between target and generated distribution, objective functions ($C_d$ and $C_g$), Norm of the gradients($D_{grad}$ and $G_{grad}$), mean and standard deviation of the two distribution during the training as a function of the number of epochs.All the statistics were computed using 100 data points sampled from the quantum generator with the optimized parameters after the training is completed.}
    \label{fig:sim_met_all}
\end{figure}

\end{document}